\documentclass[12pt]{iopart}
\usepackage{graphicx}
\begin{document}

\vspace{10pt}
\title{Fractional Quantum Hall plateaus in mosaic-like conductors}

%
%
%
%
%
%
\newcommand{\newscale}{0.2}
\newcommand{\newscaleM}{0.05}
\newcommand{\newscalex}{0.375}
\newcommand{\newscaley}{0.05}




\author{Ferdinand Kisslinger$^1$, Dennis Rienm\"uller$^1$, Christian Ott$^1$,Erik Kampert$^2$, Heiko B. Weber$^1$}
\address{$^1$ Lehrstuhl f\"ur Angewandte Physik,
Friedrich-Alexander-Universit\"at Erlangen-N\"urnberg (FAU), Staudtstr.\ 7,
91058 Erlangen, Germany}
\address{$^2$ Dresden High Magnetic Field Laboratory,
Helmholtz-Zentrum Dresden-Rossendorf, Bautzner Landstra\ss e\ 400,
01328 Dresden, Germany}

\ead{heiko.weber@fau.de}

\begin{abstract}
We report a simple route to generate magnetotransport data that results in fractional quantum Hall plateaus in the conductance.  Ingredients to the generating model are conducting tiles with integer quantum Hall effect and metallic linkers, further Kirchhoff rules. When connecting few identical tiles in a mosaic, fractional steps occur in the conductance values. Richer spectra representing several  fractions occur when the tiles are parametrically varied. Parts of the simulation data are supported with purposefully designed graphene mosaics in high magnetic fields.  The findings emphasize that the occurrence of fractional conductance values, in particular in two-terminal measurements, does not necessarily indicate interaction-driven physics. We underscore the importance of an independent determination of charge densities and critically discuss similarities with and differences to the fractional quantum Hall effect.
\end{abstract}
\maketitle





\date{\today}


\section{Introduction}

The observation of plateaus in the low-temperature Hall-conductance of a two-dimensional conductor in conjunction with a vanishing longitudinal resistance is the transport signature of quantum Hall effect (QHE)\cite{Klitzing1986}. Within the non-interacting electron picture, the plateaus are expected to appear at integer Landau Level (LL) filling fractions $\nu$ with a (Hall-)conductance of $\nu e^2/h$, as found with high precision in many two-dimensional electron gases. In some materials, at even lower temperatures, a substructure in the (Hall-)conductance with additional plateaus appears, with $\nu$ given by fractional values $f$ , which seem to follow $f=p/q$ with $p={1,2,3,\ldots}$ and $q={3,5,7,\ldots}$ \cite{Stormer1983}. This phenomenon is commonly termed fractional quantum Hall effect (FQHE) and explained by collective states of matter that are stabilized at fractional filling factors \cite{Stormer1999}.\\ 

The formation of LLs and integer QHE appears similarly in graphene which we chose in this manuscript as a representative of a 2D conductor. The QHE of monolayer graphene differs from conventional two-dimensional conductors as it shows half-integer values of $4e^2/h$ in the conductance:
\begin{eqnarray}
G_n = \frac{4e^2}{h} \left( n + \frac{1}{2} \right) \qquad \texttt{with} \quad n=0,\pm1,\pm2,\ldots \label{eqn:MLG_sequence}
\end{eqnarray}
This particularity origins in its band structure that is often related to massless Dirac fermions. In monolayer graphene, the lowest LL is shared equally by electrons and holes. The LL energy is given by $E_{LL}=\sqrt{2ev_F^2\hbar B n}$ \cite{Novoselov2005}. Apart from this remarkable difference in the LL spectra and prefactors for the conductance plateaus, QHE in graphene is very similar to standard 2D conductors and can be described by a non-interacting electron picture.

In the early days of graphene research, fractional values of $e^2/h$ in the two-terminal conductance were interpreted as heralding FQHE in graphene \cite{Du2009,Bolotin2009}. Similarly, rich spectral features have been found in bilayer graphene \cite{Bao2010}. Since then, a variety of detailed investigations of FQHE states in monolayer \cite{Dean2011,Ghahari2011,Feldman2013} and bilayer graphene \cite{Ki2014,Kou2014,Maher2014} were reported.

Another particularity of 2D conductors is that extended, i.e. one-dimensional line defects, play a much more important role than in 3D conductors. Whereas in 3D conductors, electrons can always propagate as weakly perturbed Bloch waves around a line defect, in 2D the line defect intersects the plane and thus provides a subdivision into a mosaic-like conductor. Recent studies elucidated how stacking faults can drastically alter the electronic properties of bilayer graphene \cite{Alden2013,Butz2014,Kis2015,Ju2015,Shallcross2016,Yin2016}. In particular, partial dislocations split the area into a mosaic consisting of two types of regions which exhibit quantum mechanically orthogonal wavefunctions \cite{San-Jose,Kisslinger2017b}. This decomposition is not limited to atomically thin materials: for embedded 2D semiconductor heterostructures, substrate steps or stacking faults in the surrounding crystal can create such extended defects. 


Hence, when considering a real material, line defects and the resulting mosaic-like tiling are a barely avoidable fact in 2D conductors, if the sample is large enough. An indicator for the density of line defects is the charge carrier mobility $\mu$. High values of $\mu$, i.e. a long mean free path, indicate a small number of line defects but not necessarily zero.

As an approach to understand the impact of mosaic-like decomposition of QHE materials, we assume intact tiles, which may differ in shape, but have well-defined QHE properties. In particular, they have a spectral behavior like in eq. \ref{eqn:MLG_sequence}, and edge states are formed. We further treat the interaction in between the tiles as classical, i.e. parametrized only by potential difference and electrical current. This classical treatment is motivated by the assumption that if there is any difference between 
adjacent edge states, whether its origin is spectral, topological or expressed by other differing quantum numbers, phase information is washed out along the 1D extended intersection.

In this manuscript, we consider the impact of subdivision of a 2D conductor in the QHE regime using two techniques, simulations and transport experiments on artificially subdivided graphene samples. We present a simple scheme that leads to a phenomenology with remarkable similarity to fractional quantum Hall phenomena without involving novel ground states. Similarities and differences are critically discussed, elaborating a thought-provoking contribution to the field.

\begin{figure}[hbtp!]
    \centering
        \includegraphics[width=87mm]{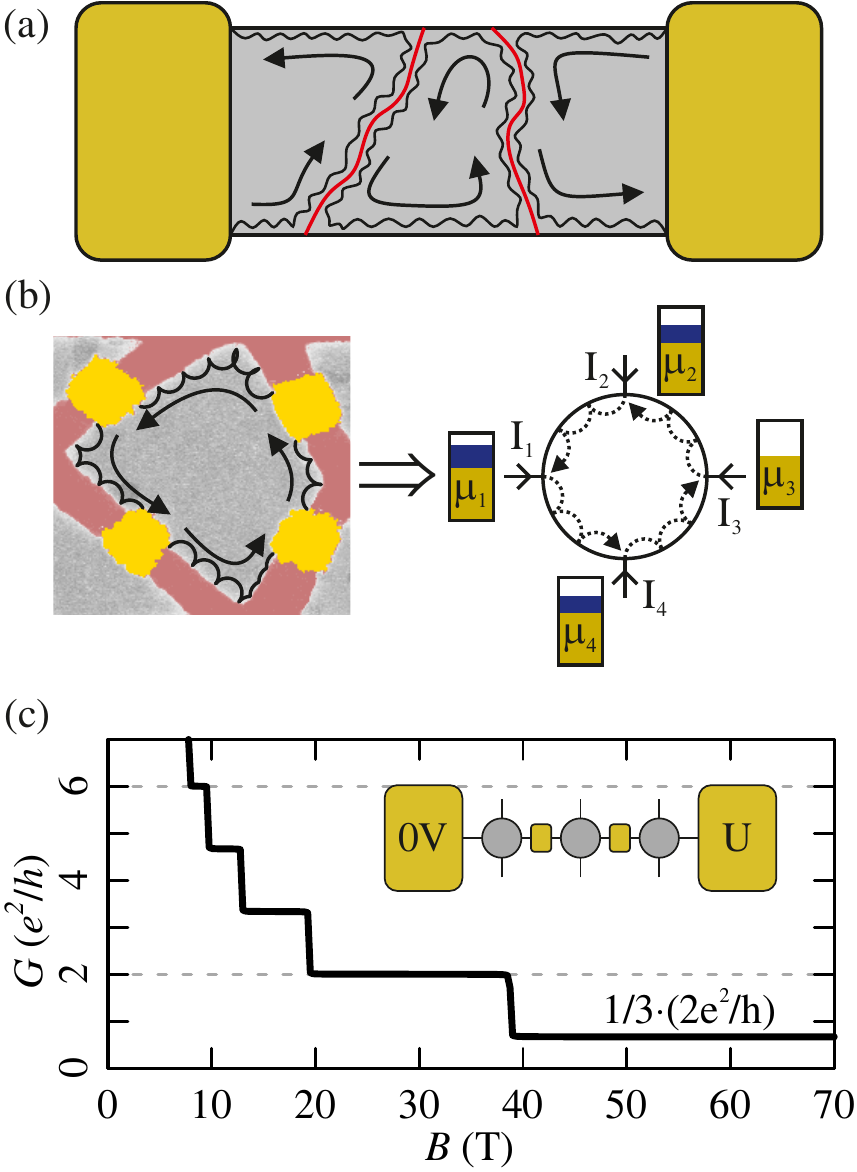}
                        \caption{(a) Sketch of edge states in a 2D sample (grey) contacted with two metal electrodes (yellow). The intersection of sample by 1D defects (red lines) leads to decomposition in three regions. (b) The situation in (a) is modelled by a network of homogeneous conductors, which is connected via metallic linkers to the neighbouring regions. The transport is described by edge currents and the potential of metallic linkers. (c) Magnetoconductance of the network corresponding to (a). Its implementation is displayed as inset.}
    \label{fig:1}
\end{figure}

\section{Model and Experiment}
We assume a geometry in which a 2D conductor is composed of unperturbed and homogeneous areas (tiles) that are in the QHE regime (see Fig. \ref{fig:1}\,a). The electrical connection, both in model and experiment, is made by metallic interfaces that can transmit a current from one tile to its neighbour. The metallic character of this linking unit ensures that edge states from one tile can not communicate quantum mechanically with edge states from the neighbour tile. This classical communication simplifies to a defined potential and an ohmic current through the connecting unit (Fig. \ref{fig:1}\,b). Experimentally, this model is realized by graphene segments, interconnected with metallic stripes. Theoretically, each tile and its linkers are represented by textbook Landau-B\"uttiker modelling of quantum coherent QHE samples with ideal metallic leads \cite{Datta}, a concept that received additional justification by topological considerations \cite{Haldane1988}. The mosaic material is then considered as a network of such tiles, for which classical Kirchhoff rules are valid.

\subsection{Model}
We model the QHE in a perfect tile with four ideal terminals by a conductance matrix (magnetic field pointing into the plane) \cite{Buettiker1998,Datta}:   

\begin{equation}
	\left(
	\begin{array}{c}
		I_1 \\ I_2 \\ I_3 \\ I_4
	\end{array}
	\right)
		=
		\left(
	\begin{array}{cccc}
		 -G_{n}&G_{n}&0&0\\
		0 & -G_{n} & G_{n} & 0 \\
		      0 & 0 & -G_{n} &G_{n} \\
			G_{n} & 0 & 0 &-G_{n} 
	\end{array}
	\right)
	\cdot
	\left(
	\begin{array}{c}
		\mu_1 \\
		\mu_2 \\
		\mu_3 \\
		\mu_4 
	\end{array}
	\right)
\end{equation}
where the resulting electron current into the tile is defined as positive and the conductance $G_n$ is a multiple of the conductance quantum. The multiplicity is given by the integer number of occupied edge states (the number of LL below the Fermi energy $E_F=\hbar v_F \sqrt{\pi n_s}$, with $n_s$ the charge carrier density). Thus, $G_n$, which essentially reflects the contact resistance of current-carrying leads, is given by eq. (\ref{eqn:MLG_sequence}) for monolayer graphene, with all edge states occupied below the Fermi energy. 

Such tiles are connected to build a four-terminal resistor network, implemented as a system of linear equations that represents the Kirchhoff rules and boundary conditions for currents and voltages, which is solved numerically using a sparse matrix algorithm \cite{Kisslinger2017}. 

The model is suited to treat regular tilings (matrix arrangements) as well as irregular mosaics (see for example Fig.\,\ref{fig:2}). In the latter, we limited the treatment to four terminals per tile or less, a generalization to higher numbers is straightforward. Note that, within a B\"uttiker treatment of QHE, the transport is independent of the geometrical shape of the tile because charge is carried by edge states. The model is only valid in the low temperature/high field limit, in which the single tiles show QHE.


\subsection{Experimental methods}
Monolayer graphene (MLG) was grown on the silicon terminated face SiC(0001) of commercial wafers (Cree, 4H-SiC HPSI, nominally on-axis) at $\sim 1700\,\mathrm{C^{\circ}}$ and $\sim 900\,\mathrm{mbar}$ Ar pressure for 20-30\,min. The growth process results in an average coverage with slightly more than one monolayer to obtain a conducting sheet \cite{Emtsev2009}. Resistor networks were patterned using  standard e-beam lithography followed by either evaporation of Ti/Au contacts or O$_2$-plasma etching to remove unwanted graphene areas. Electrical magneto-transport measurements were carried out at the Dresden High Magnetic Field Laboratory in pulsed fields (pulse duration of about 150\,ms) up to $\sim 70\,$T at $T\sim 1.4\,$K, to reach the quantum regime in the highly n-doped ($\sim 5\cdot10^{12}/\mathrm{cm^2}$) epitaxial graphene samples \cite{Jobst2010}.

\section{Results}
\subsection{Simulations: Mosaics with identical tiles}
The essence how tiles generate fractions of quantum conductance becomes clear in a Gedankenexperiment: when a sequence of three identical tiles form the sample that is contacted by two current leads (see Fig. \ref{fig:1}\,c), the conductance $G_{total}$ equals $(\sum_{i=1}^{3}{1/G_{n,i}})^{-1}$, delivering a prefactor of $1/3$ as compared to equation \ref{eqn:MLG_sequence}. This simple example teaches that in a two-terminal experiment the subdivision easily generates a fraction of quantum conductance within the given LL sequence. In particular for LL index $n=1$ the two-terminal conductance $G_{total}$ reaches $2e^2/h$ that is the expectation of an untiled QHE sample at $n=0$ LL occupied. Such a voltage division has been discussed in the early days of QHE, where it has supported the notion of edge channel transport \cite{Muller1990}, and similarly in graphene \textit{p-n-p} junctions \cite{Oezyilmaz2007}.

In a two-dimensional mosaic, the current inflowing into an arbitrary tile may be distributed over the other metallic linkers. When sticking to two-terminal conductance measurements of a mosaic, more complicated fractions of the conductance quantum $e^2/h$ occur. For identical tiles, the mosaic acts as a voltage divider, generating fractional prefactors to the quantum conductance. When the magnetic field $B$ is swept, or alternatively a gate voltage tunes the Fermi level through the LL sequence, the conductance of each tile changes simultaneously and the fractional prefactors are unaffected.

\begin{figure}[hbtp!]
    \centering
        \includegraphics[width=87mm]{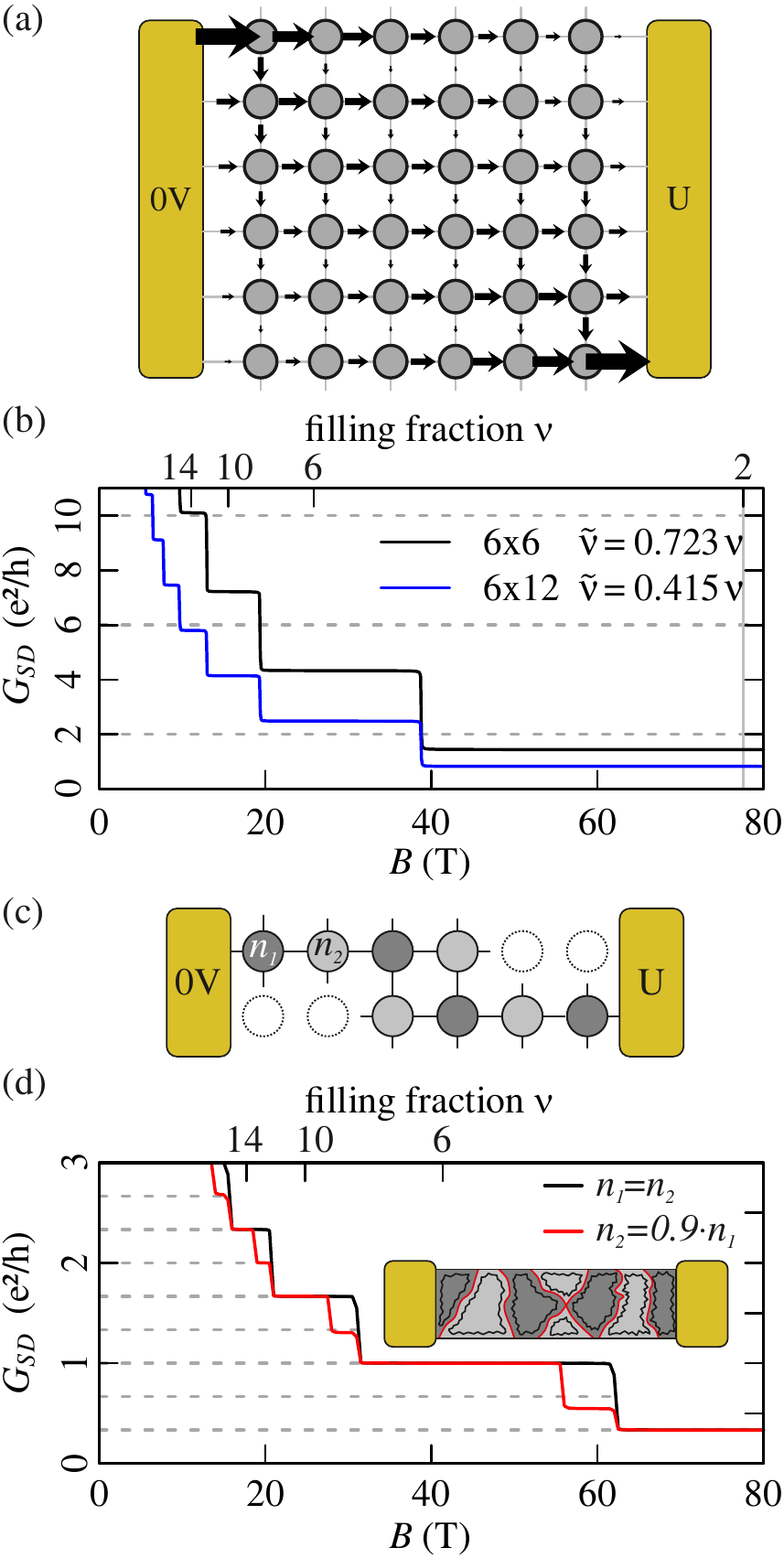}
                        \caption{(a) Current distribution in a rectangular $6 \times 6$ matrix at $50\,$T (charge carrier density $n_s=3.75\cdot 10^{12}/\mathrm{cm^2}$). Comparison of magnetoconductance of the $6 \times 6$ and a $6 \times 12$ matrix. The top axis indicates the filling fraction calculated via $\nu=hn_s/eB$. (c) Sketch of bipartite network with $n_1=6\cdot 10^{16}/\mathrm{cm^2}$ and $n_2=0.9\cdot n_2$. (d) Magnetoconductance of the network in (c). Inset: illustration of sample, which could correspond to this network.}
    \label{fig:2}
\end{figure}

We stick to simple geometries and assume a regular array in which identical tiles are arranged as a square $N \times N$ matrix. Classically, in the absence of magnetic fields, the resistance of a large square $R_{\opensquare}$ is the same as the resistance of the tile itself, because the effect of resistances in series and resistances in parallel exactly cancel each other. In the square and regular quantum resistor network, the effect of sequential voltage drop by a factor of $N$ does not compensate with the effect of parallel arrangement of $N$ channels, because quantum Hall currents flow perpendicular to the rows. A simulation of the current pathways in a $6\times 6$ matrix is shown in Fig. \ref{fig:2}\,a. Fig. \ref{fig:2}\,b shows the simulated conductance of this square network. While the step sequence is identical to the step sequence of the individual tile, the conductance values at the steps give \textit{apparent filling fractions} $\tilde{\nu}$ of value $\approx 0.723\,\nu$, behind which we expect a rational number. In the limit of large $N$, the apparent filling fraction $\tilde{\nu}$ approaches $\nu/\sqrt{2}$. An interesting case is now to elongate the matrix and compare the square $N \times N$ with $N \times 2N$. The result is displayed in Fig. \ref{fig:2}\,b. The plateau height is not simply reduced by a factor of two, but is only reduced to a value close to $1.74$.

It is not surprising that when the matrix $N \times M$ is further varied many values can be generated. We now study elongated matrices with low $N$ and $M$, similar to standard Hall bars with few defect lines.


For the consideration of more realistic, i.e. irregularly arranged mosaics, we choose an incomplete $2 \times 6$ matrix, displayed in Fig. \ref{fig:2}\,c, purposefully designed to display FQHE-typical features.  The result of the simulation is shown in Fig. \ref{fig:2}\,d. It can be seen that if we choose an input value of $\nu=2$ (final plateau), the voltage divider mechanism results in an apparent filling fraction $\tilde{\nu}=1/3$. A second important finding of this example is that an apparent filling fraction of $\tilde{\nu}=1$ exists at lower fields, at which the true filling factor is $\nu=6$.  Up to this point, we considered two-terminal measurements and identical tiles. In a next step, we will present, how (i) releasing the constraint of identical tiles and (ii) four-terminal measurements affect the phenomenology.

\subsection {Simulations: Mosaics with different tiles}

We consider the case that a 2D mosaic has sharply defined domains of slightly varying charge density $n_1, n_2,\ldots$. For atomically thin materials as graphene, this may result from adsorbat puddles, topgate non-uniformities, substrate inhomogeneities etc. For the more traditional 2DEGs in semiconductor heterojunctions, a variation in the crystalline environment, stacking faults, step-like inhomogeneities could define domains with different induced charge densities.

We model the influence of this mosaic-like decomposition in the LL regime by the matrix approach sketched above with the straightforward modification that different charge densities $n_i$ are assigned to the tiles. This goes along with different LL stacks in neighboring tiles. For simplicity, we chose a bimodal distribution of $n_1$ and $n_2$ for two sorts of tiles, resulting in a two-color mosaic (cf. Fig. \ref{fig:2}\,c). As argued above, we assume that within one domain, the QHE is nicely described by edge states. The interaction with its spectrally different neighbor along a sharp separation line is assumed as classical, because any phase-dependent effect of the spectrally different edge states would average out along the line.

In Fig. \ref{fig:2}\,d we display results obtained with such a bimodal mosaic, with charge densities differing as little as $10\%$ ($n_1 = 0.9\,n_2$). The influence on the two-terminal conductance is remarkable: each step splits into two steps, and new plateaus at fractional apparent filling factors occur. The origin can easily be understood: with increasing field, charge transport through LL states with higher index are subsequently switched off, this happens at different fields, each time reducing the conductance through the sample. Note that for the specific geometry chosen, apparent filling factors (4/3 etc.) can be read off. 
The decomposition here resembles experiments with electrostatically gated tiles of different charge densities in graphene \textit{p}-\textit{n} boundaries, which show unconventional apparent filling fractions \cite{Williams2007,Oezyilmaz2007,Abanin2007}. 

We conclude that by assuming sharply defined charge density domains, which form classically behaving interface lines, our model can straightforwardly generate rich spectral features in the magnetoconductance.

\begin{figure*}[hbtp!]
    \centering
        \includegraphics[width=180mm]{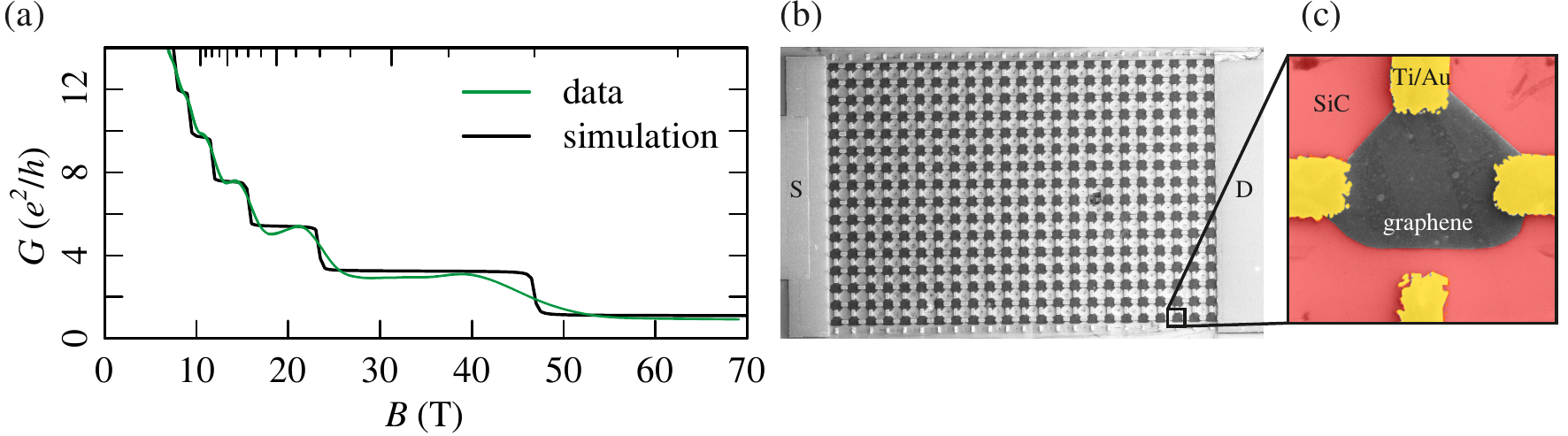}
                        \caption{(a) Magnetoconductance of an experimental $16 \times 23$ network of epitaxial monolayer graphene with Ti/Au linkers (green line). The measurement was performed in pulsed magnetic fields. The simulation data (black line) is calculated using the charge carrier density $n_s=4.5\cdot 10^{12}/\mathrm{cm^2}$ obtained from Shubnikov-de Haas oscillations at a test structure on the very same chip. (b) Scanning electron micrograph of the sample. (c) Close-up of a single tile.}
    \label{fig:3}
\end{figure*}

\subsection{Experiments: mosaic with identical tiles}

Our interest is to understand charge transport in the quantum regime of 2D conductors, which often disintegrate into tiles of a mosaic, separated by barely visible domain boundaries \cite{Alden2013,Butz2014}. Again, we assume the latter to behave classically. In order to translate the model into a real device, we opted for artificial disintegration of a graphene sheet into tiles, which we connected via metallic strips. Fig. \ref{fig:3}\,b,\,c show an example of such an experimental realization: a matrix arrangement of graphene sheets ($16 \times 23$), interconnected by four metallic terminals each (except at the edge), which is electrically contacted by two electrode banks at the left and right side, respectively (source and drain). Epitaxial graphene displays QHE and has rather homogeneous charge density $4.5\cdot10^{12}/\mathrm{cm^2}$, however, the devices are not fully identical because part of the area is actually bilayer graphene ($\sim 20\,$\%). The two-terminal conductance measurement has been carried out at low temperatures and high magnetic fields, the results are displayed in Fig. \ref{fig:3}\,a, together with simulations performed for this network configuration. Obviously, the LL structure expresses as steps in the conductance, very similar to standard QHE but the step heights are reduced by a prefactor that has the same origin as the square matrix in Fig. \ref{fig:2}\,a and b. This emphasizes how easy deviations from standard conductance values (cf. eqn. (\ref{eqn:MLG_sequence})) can be generated. Obviously, in our experiment the steps occur rounded which can be traced back to inhomogeneities (presumably intra-tile as well as inter-tiles). Despite we heralded multiple steps and rich spectral features as a consequence of inhomogeneities, in the present experiment the large number of tiles as well as the insufficiency of accurately defined tile parameters mask the occurrence of additional steps. 

The findings underscore the validity of the simulations that in a mosaic fractions of $e^2/h$ conductance plateaus can easily be generated. They further make clear that rich spectral features can only be expected when few tiles and sharply defined tile properties are present. 

\begin{figure*}[hbtp!]
    \centering
        \includegraphics[width=180mm]{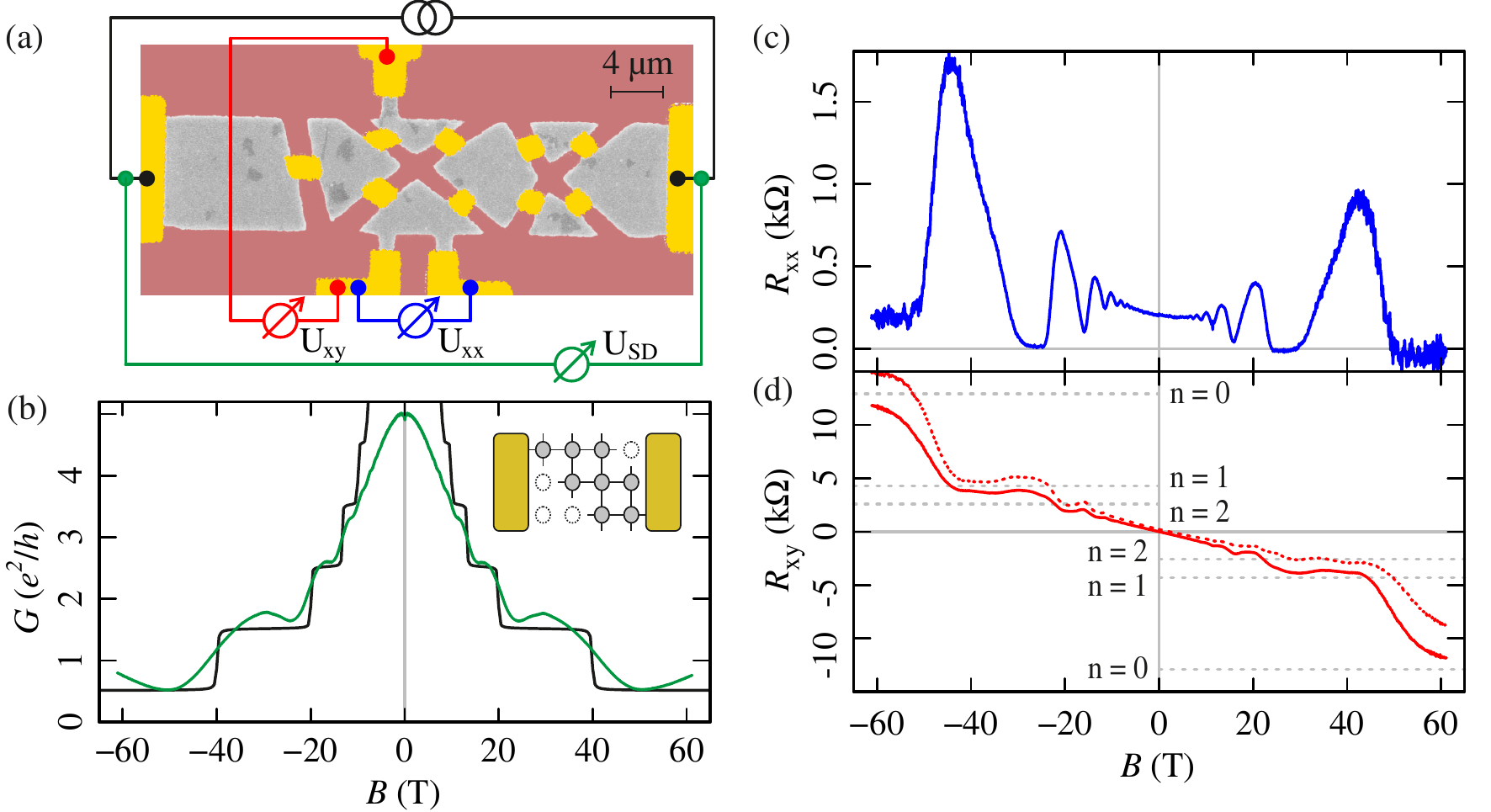}
                        \caption{(a) Colored scanning electron micrograph of a purposefully designed monolayer graphene sample (color code: monolayer$\,=\,$light gray, bilayer$\,=\,$darker gray, Ti/Au linkers$\,=\,$yellow, substrate$\,=\,$red). Current source and voltage probes are sketched. (b-d) Transport data recorded in pulsed magnetic fields. Color code refers to (a). (b) Two-terminal magnetoconductance. Experiment is shown as green line and simulation as black line. The charge carrier density $n_s=3.86\cdot 10^{12}/\mathrm{cm^2}$ for simulation was obtained from Shubnikov-de Haas oscillations at a test structure on the very same chip. Inset: Sketch of simulated device corresponding to (a). (c) Longitudinal resistance measured at a single tile. (d) Transverse resistance measured between different tiles. Raw data is shown as dashed line and the anti-symmetrized data as solid line.}
    \label{fig:4}
\end{figure*}

In that spirit, we applied the same experimental scheme to a sample that consists of only eight tiles, again interconnected by metal strips (see Fig. \ref{fig:4}\,a). Obviously, the number of terminal per tile varies from two to four. We selected the tiles such that monolayer areas (in light gray) dominate the edges of the tiles. The undesired bilayer areas that appear darker in the SEM micrograph are to a high extent isolated within the tiles. Whether or not they reach the edge at some locations cannot fully be resolved from the micrograph. The corresponding two-terminal measurements from source to drain shows features similar to QHE, but the plateaus are not ideally developed and show an overshooting conductance between the plateaus that may be related to short-channel effects \cite{Abanin2008}. In order to link up with the simulations, we mapped the sample onto the matrix scheme (see inset in Fig. \ref{fig:4}\,b), assuming identical tiles, and calculated the corresponding magnetoconductance (Fig. \ref{fig:4}\,b). Again, the occurrence of apparent filling fractions $\tilde{\nu} < \nu$ can be read off.

\subsection{Four-terminal analysis}

This sample further allows to read out four-terminal measurements (see Fig. \ref{fig:4}\,a), which are of great importance in QHE and FQHE physics. The simultaneous occurrence of plateaus in $R_{xy}$ and a vanishing $R_{xx}$ are considered as indicative for QHE. For FQHE, the criteria are weakened: fractional filling factors are read off from $R_{xy}$, but $R_{xx}$ does not necessarily vanish. For the sample displayed in Fig. \ref{fig:4}\,a, $R_{xx}$ shows Shubnikov-de Haas oscillations which indeed vanish at specific fields (see Fig. \ref{fig:4}\,c). It is important to stress that $R_{xx}$ vanishes \textit{only if} the two voltage probes contact the very same tile. When measuring $R_{xy}$, which is probed at different tiles, the experiment delivers a slightly asymmetric curve with plateau like features, which is plotted in Fig. \ref{fig:4}\,d as dashed line. We symmetrize the data with respect to $B$ inversion; the result is displayed as solid line. Remarkably, this four-probe $R_{xy}$ shows conductance plateaus at usual monolayer graphene values with no modification as compared to eqn. (\ref{eqn:MLG_sequence}). This experimental result goes fully along with corresponding simulations. We conclude: We find no conditions where $R_{xx}$ vanishes while $R_{xy}$ displays fractional conductance plateaus in a four-terminal measurement.

\subsection{Critical discussion}

The introduction of mosaic-like patterns in the description of a 2D conductor in the QHE regime introduces a voltage divider, which trivially generates fractions of conductance when the tiles are arranged in series. Within our multiterminal simulation that links QHE tiles via Kirchhoff network rules, all conductances taken at any points will have  fractional values of the quantum conductance for finite systems. 

This similarity demands for a critical comparison with FQHE found in many different 2D materials, which, however, is interpreted as an interaction driven electronic low temperature phase. According to our findings, two-terminal measurements that display fractional conductance plateaus are prone to competing interpretations (in terms of mosaic/FQHE), while the higher information content in four terminal measurements should be suited for an unambiguous assignment, provided that the charge density is known. A detailed knowledge on this latter quantity would resolve the ambiguity of filling fraction and apparent filling fraction.  We should emphasize that the mosaic model in its current stage would equally predict nearly arbitrary fractions (e.g. 1/4 as well as 1/3). This is certainly another difference to the phenomenology in FQHE research that favors certain fractions, while e.g. 1/4 is not reported. Also the hierarchy of states in FQHE \cite{Jain} is not found within our scheme.

Nevertheless, we suspect that the mosaic-like interpretation becomes visible in experiments. One example is the electron transport through graphene in which \textit{p-n} junctions have been created by external gates. The \textit{p} and the \textit{n} region in which hole and electron like LLs are formed have a clearcut interface at which the countercirculating edge states classically communicate very similar to our model, a fact that has been termed \textit{equilibration} area \cite{Williams2007}. As a consequence, fractional values of conductance and Fano factors have been experimentally demonstrated \cite{Williams2007,Abanin2007}. 

Another important example, in our opinion, is bilayer graphene. While in monolayer graphene interaction-driven effects are sparse, in the bilayer rich spectral features have been assigned to the formation of many-body effects, among which insulating phase \cite{Weitz2010,Velasco2012}, FQHE and further exotic phases are most prominent \cite{Ki2014,Kou2014,Maher2014}. It was shown, however, that one of the qualitative differences between mono- and bilayer are stacking faults that are trivially absent in monolayers and ubiquitous in bilayers. Stacking in graphene bilayers follows the Bernal stacking scheme, which has a two-fold degeneracy (often termed AB and AC stacking). In a real sample, there are areas that are AB stacked, other areas are AC stacked within the same sheet.  They are interfaced by partial dislocations (or structural solitons) that have nicely been imaged and characterized \cite{Alden2013,Butz2014}. Partial dislocations are extended line defects which thread the 2D plane. They invoke a quantum mechanical separation of electronic states on both sides of the partial dislocations which can be expressed by orthogonality of eigenstates in AB and AC areas \cite{San-Jose,Kisslinger2017b} close to charge neutrality. Other groups have chosen a description in terms of topological properties of the domains, and topological edge states that run along the partial dislocation lines \cite{Ju2015,Yin2016}. The latter picture immediately links up to the edge state description of QHE \cite{Datta,Haldane1988}. Altogether, we face a situation where two decoupled edge states (characterized by different quantum numbers) run along the two sides of a 1D interface. This motivates that the electronic communication between two neighboring tiles can be modeled classically (see above). We propose that the rich spectral features that have been observed in suspended bilayer samples \cite{Du2009,Kou2014,Maher2014}, including fractional conductance values and unusually high condensation energies \cite{Skachko2010}, should be carefully reconsidered in terms of mosaic conductors. Note also that partial dislocations provide a good explanation of the coexistence of insulating/conducting properties of bilayers, previously understood as `insulating phase' \cite{Shallcross2016}.

\begin{figure}[hbtp!]
    \centering
        \includegraphics[width=87mm]{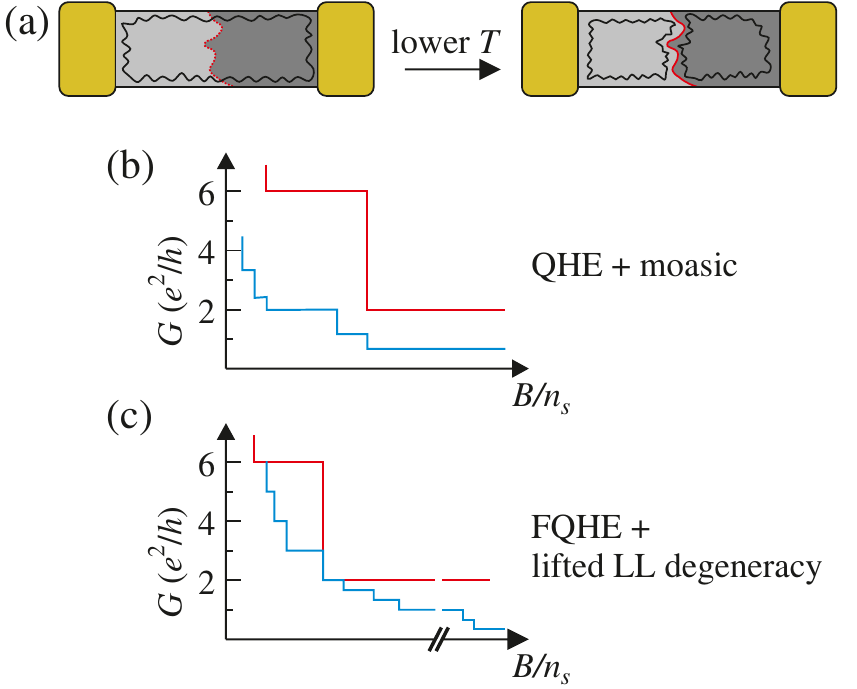}
                        \caption{Gedankenexperiment: (a) Phenomenological representation of a sample which decomposes in two areas with well-separated edge states when thermal energy becomes lower than a specific threshold energy. The threshold energy will depend on the mechanism which causes the decomposition. (b) and (c) Comparison between fractional conductance values as a function of $B/n_s$ expected in a mosaic of QHE conductors (b) and in FQHE regime (c). The regular monolayer graphene QHE (red line) is indicated as reference.}
    \label{fig:5}
\end{figure}

Within our model, a mosaic of identical charge densities would not generate additional conductance steps. Once there is a charge density contrast between the tiles, expressed by different values $n_i$, additional steps with fractional distances occur in the conductance (cf. Fig. \ref{fig:2}\,c,\,d). This directs our attention to subtle mosaic-like charge density patterns that may occur by subtle differences of the electrostatic environment of the 2D conductor. The expression \textit{mosaic like} emphasizes our notion that there are sharp one-dimensional interfaces between homogeneous areas. In a Gedankenexperiment, we consider a high-mobility sample with regions of slightly different charge densities $n_1$ and $n_2$, a difference so small that it can barely be detected at room temperature by Hall measurements. When lowering the temperature, the electronic system forms LL spectra with a slight spectral mismatch expressed by an energy difference $\Delta E$ and still appears homogeneous as long as ($\Delta E < k_B T$). In this case we expect a regular integer QHE. When the temperature is further lowered ($\Delta E > k_B T$), the spectral mismatch may electronically decouple the two neighboring areas. The resulting decomposition of the electronic system when the temperature is lowered is schematically depicted in Fig. \ref{fig:5}\,a: The spectrally different areas may develop their own edge state loops. If there was such a decomposition as a function of temperature, it would presumably result in a lowering of the quantum Hall conductances towards fractional values, according to our network model. It seems, however, that the QHE is robust with respect to small perturbations. A decomposition into a mosaic is only expected when quantum numbers of adjacent tiles (for example their Chern numbers)  differ. The latter scenario of a mosaic with charge density contrast would presumably result in thermally activated behavior when the temperature is swept, again a qualitative similarity to FQHE. This would similarly apply also to transport barriers that may be introduced by 1D defect lines.

While we listed many arguments that raise concern whether our model has something to do with the observations of FQHE at all, the similarities in phenomenology are surprising, all the more since this manuscript provides only the simplest conceivable approach to a network of QHE domains. Interesting routes are e.g. a proper consideration of threefold symmetry of many FHQE materials. This affects not only the electronic structure, it leads also to three equivalent but different Burgers vectors of (partial) dislocations \cite{Alden2013,Butz2014}.

\section{Conclusions}

Two-dimensional conductors are prone to contain line defects that intersect the sample into domains. We introduce a model that connects 2D conductors in the quantum Hall regime via metallic linkers. The model thus represents a 2D \textit{mosaic} conductor. The metallic linkers simulate an unspecified classical interaction between the tiles which is essentially modeled by current and electrostatic potential. The model inherently generates fractional values of conductance by a simple voltage divider mechanism. Experiments using graphene as 2D conducting tiles and gold as metallic linkers confirm the outcome of the simulations.

In particular for the case of few-tile mosaics with slightly different charge densities, rich spectral features and fractional conductance plateaus can be generated that resemble on first glance fractional quantum Hall phenomena. In a well-controlled experiment, in which the charge density is independently determined, the findings can properly be discriminated from the findings assigned to FQHE. In particular in two-terminal conductance measurements, however, the appearance of fractional conductance values are not conclusive for assigning interaction-driven physics.

The model we present can not describe the complete picture of FQHE, but given its simplicity, it comes remarkably close. It provides simple access to some generic findings like fractional values of conductances, and rich spectral features in  2D conductors. It can not explain the importance of fractional filling of Landau levels, the predominant occurrence of odd denominators, the hierarchy of states, etc.

Despite the clear differences between the phenomenology reported in FQHE literature and in our simple model, we feel an esthetical discomfort: how can the interaction-driven many-body FQHE approach on the one hand and the QHE in a mosaic on the other hand coexist and generate such similar phenomena, although they are conceptually unrelated?

\vspace*{1em}
\ack
The work was carried out in the framework of the SFB 953 and SPP 1459 of the Deutsche Forschungsgemeinschaft (DFG). We acknowledge the support of the HLD at HZDR, a member of the European Magnetic Field Laboratory (EMFL).

\section*{References}

\bibliographystyle{unsrt}

\bibliography{bibliography-fqhe_phenomena}

\end{document}